\documentclass[aps,prd,preprint,groupedaddress]{revtex4-1}

\usepackage[utf8x]{inputenc}
\usepackage{amsmath,amssymb}
\usepackage{hyperref}
\usepackage{physics}
\usepackage{graphicx}

\begin{document}


\title{Rotating Spacetimes generalizing Lifshitz Black Holes}


\author{Alfredo Herrera-Aguilar}
\email[]{aherrera@ifuap.buap.mx}
\author{Jhony A. Herrera-Mendoza}
\email[]{jherrera@ifuap.buap.mx}

\author{Daniel F. Higuita-Borja}
\email[]{dhiguita@ifuap.buap.mx}

\affiliation{Instituto de Física, Benemérita Universidad Autónoma de Puebla. Apdo. Postal J-48, C.P. 72570,  Puebla, México}

\date{\today}

\begin{abstract}
We present a spinning black hole solution in $d$ dimensions with a maximal number of rotation parameters in the context of the Einstein-Maxwell-Dilaton theory. An interesting feature of such a solution is that it accommodates Lifshitz black holes when the rotation parameters are set to zero. We verify the rotating nature of the black hole solution by performing the quasi-local analysis of conserved charges and defining the corresponding angular momenta. In addition, we perform the thermodynamical analysis of the black hole configuration, show that the first law of thermodynamics is completely consistent, and obtain a Smarr-like formula. We further study the thermodynamic stability of the constructed solution from a local viewpoint, by computing the associated specific heats, and from a global perspective, by using the so-called new thermodynamic geometry. We finally make some comments related to a pathology found in the causal structure of the obtained rotating black hole spacetime and compute some of its curvature invariants.
\end{abstract}


\maketitle

\section{Introduction\label{Sec:Intro}}
During last decades, the AdS/CFT correspondence \cite{Maldacena:1997re} has become a milestone in modern physics. The idea of connecting, through a holographic dictionary, gravitational aspects of a  theory in $d$ dimensions with quantum aspects of a field theory in $d-1$ dimensions seems very appealing. The overwhelming success of this approach to different problems in physics has motivated the quest to modify or even extend the correspondence from his original and mathematical well-defined setup \cite{Maldacena:1997re} to new frontiers. It is the case of Condensed Matter Physics where systems are usually represented by non-relativistic field theories, which are anisotropic and allow characteristic directions. The strongly interacting regime of such theories is usually difficult to solve and constitutes a rich ground for a Gauge/Gravity duality since the gravitational dual falls into the weakly interacting limit and it can be easily solved. To establish the correspondence it is crucial to have objects sharing symmetries and physical properties to be connected through the duality. In this work we will focus on anisotropic gravitational configurations, being an important case the Lifshitz spacetime which can be understood as a gravitational dual to non-Lorentz invariant quantum field theories \cite{Kachru:2008yh}   
\begin{equation}\label{eq:LifshitzET}
ds^2=-\left(\dfrac{r}{\ell}\right)^{2z}dt^2+\dfrac{dr^2}{\left(\dfrac{r}{\ell}\right)^2}+\left(\dfrac{r}{\ell}\right)^2d\vec{x}{}^2.
\end{equation}
The metric \eqref{eq:LifshitzET} is invariant under the anisotropic scale transformations
\begin{equation}
D_z: \,\,\,\, t\mapsto\lambda^zt,\hspace{5mm}r\mapsto\dfrac{r}{\lambda},\hspace{5mm}\vec{x}\mapsto\lambda\vec{x},
\end{equation}
with $z$ known as the dynamical critical exponent that represents the degree of anisotropy of the spacetime (the isotropic relativistic case $z=1$ corresponds to AdS). 
It is also invariant under space and time translations and spatial rotations:  
\begin{eqnarray}\label{eq:s}
H:&  & \,\,\, t \longrightarrow \,\,\,t'=t+a;\nonumber \\
P^{i}:&  & x^{i}\longrightarrow x^{'i}=x^{i}+a^{i}; \\
L^{ij}:&  & x^{i}\longrightarrow x^{'i}=L^{i}_{j}x^{j}.\nonumber 
\end{eqnarray}
To include finite temperature effects in the holographic correspondence it is necessary to introduce black hole configurations, being the first asymptotically Lifshitz black hole the one reported in \cite{Taylor:2008tg}. Since then, many static asymptotically Lifshitz Black Holes have been found in several gravitational theories \cite{AyonBeato:2009nh} and different dimensions \cite{AyonBeato:2010tm}, failing consistently to have a rotating version. The only published advance in this direction is given in the context of New Massive Gravity with $z=3$ through a boost \cite{Sarioglu:2018rhl}, with a detailed thermodynamical study in \cite{Kolev:2019tqa}. We will extend these works to arbitrary dimensions and construct a rotating configuration that accommodates Lifshitz black holes in certain limit. To achieve this aim we will organize the paper as follows: In Sec. \ref{Sec:Ansatz} we will present the Einstein-Maxwell-Dilaton theory and our black hole metric ansatz with the maximal number of independent rotation parameters in a $d$-dimensional spacetime. Once the field equations are obtained and a rotating black hole solution is found, the thermodynamical study has to be performed. Then we shall need to compute global conserved charges related to the Killing vector fields of our ansatz, different methodologies can be applied to achieve this goal. However, we will use the off-shell quasi-local formalism \cite{Kim:2013zha,Hyun:2014kfa,Hyun:2014sha,Gim:2014nba} since it is straightforward and well-defined for different kinds of asymptotic behavior of the fields. Thus, Sec. \ref{Sec:Thermodynamics} will be devoted to briefly review the quasi-local formalism and to compute the relevant thermodynamical quantities, unveiling the rotating nature of the black hole configuration by calculating its angular momenta. 
We introduce a generalization of the Hawking temperature definition that is compatible with the third law of Thermodynamics, in the sense that avoids having degenerate states with finite entropy and vanishing temperature, and allows for a reasonable description of phenomenological condensed matter systems that attain a phase transition at finite temperature \cite{Hartnoll2016vb,Sachdev2012qw}. In Sec. \ref{Sec:NTG} we use these thermodynamical quantities to construct a geometrization of the thermodynamic space on the basis of the Gibbs free energy and analyze possible phase transitions based on the specific heat and curvature invariants. We then engage with
the spacetime causal structure in Sec. \ref{Sec:Physics}. We will end with some conclusions and remarks
regarding the scope of our work in Sec. \ref{Sec:Conclusions}.

\section{The Gravitational Configuration \label{Sec:Ansatz}}
The first Lifshitz black hole solution \cite{Taylor:2008tg}
\begin{equation}\label{eq:Ansatz}
ds^2=-\left(\dfrac{r}{\ell}\right)^{2z}f(r)dt^2+\dfrac{dr^2}{\left(\dfrac{r}{\ell}\right)^2f(r)}+\left(\dfrac{r}{\ell}\right)^2d\vec{x}{}^2, 
\end{equation}
was engineered in the context of the Einstein-Maxwell-Dilaton (EMD) theory defined by the following action  
\begin{equation}\label{eq:EMD}
S[g^{\mu\nu},\phi,A_{\mu}]=\int{d^dx\sqrt{-g}\left(\dfrac{R-2\lambda}{2\kappa}-\dfrac{1}{2}\nabla_\mu\phi\nabla^\mu\phi-\dfrac{1}{4}e^{-b\phi}F_{\mu\nu}F^{\mu\nu}\right)},
\end{equation}
with the following field equations
\begin{subequations}\label{eq:EoM}
\begin{align}
\delta_gS & =0\Longrightarrow G_{\mu\nu}+\lambda g_{\mu\nu}=\kappa T_{\mu\nu},\\
\delta_AS & =0\Longrightarrow \partial_\mu\left(\sqrt{-g}e^{-b\phi}F^{\mu\nu}\right)=0,\\
\delta_\phi S & =0\Longrightarrow \Box\phi=-\dfrac{b}{4}e^{-b\phi}F_{\mu\nu}F^{\mu\nu}, 
\end{align}
\end{subequations}
where 
\begin{equation}\label{eq:EnergyMomentum}
T_{\mu\nu}=\nabla_\mu\phi\nabla_\nu\phi-\dfrac{1}{2}g_{\mu\nu}\nabla_\alpha\phi\nabla^\alpha\phi+e^{-b\phi}\left(F_{\mu\alpha}{F_{\nu}}^\alpha-\dfrac{1}{4}g_{\mu\nu}F_{\alpha\beta}F^{\alpha\beta}\right). 
\end{equation}
It would be interesting to find a rotating version of this family of Lifshitz black holes. In order to accomplish this aim, inspired by the (A)dS rotating solutions found in \cite{Awad:2002cz}, we will make use of the following metric ansatz 
\begin{align}\label{eq:Cov_Ansatz}
ds^2=-\left(\dfrac{r}{\ell}\right)^{2z}f(r)\left(\Xi dt-\sum_{i=1}^{n}a_id\phi_i\right)^2 & +\dfrac{r^2}{\ell^4}\sum_{i=1}^{n}\left(a_i dt-\Xi \ell^2 d\phi_i\right)^2+\dfrac{dr^2}{\left(\dfrac{r}{\ell}\right)^2f(r)}\nonumber\\
 & -\dfrac{r^2}{\ell^2}\sum_{i<j}^{n}\left(a_id\phi_j-a_jd\phi_i\right)^2+\left(\dfrac{r}{\ell}\right)^2d\vec{y}{}^2, 
\end{align}
where $n=[(d-1)/2]$ is the maximal number of independent rotation planes in $d$ dimensions, each one related with a rotation parameter $a_i$, $\Xi=\sqrt{1+\sum_{i=1}^{n}a_i^2/\ell^2}$, and $d\vec{y}{}^2=dy^kdy^k$ is the Euclidean metric on a $(d-2-n)$-dimensional submanifold. We will prove later on that such a configuration has indeed non-trivial angular momenta. In addition we shall assume that matter fields share the same spacetime symmetries, taking the following scalar field and vector potential ans\"atze \cite{Awad:2002cz}   
\begin{equation}\label{eq:FieldAnsatz}
\phi=\phi(r), \qquad A=-\Xi h(r)dt+\sum_{i=1}^n a_i h(r)d\phi_i.
\end{equation}
By collecting all this information into the field equations \eqref{eq:EoM} we arrive to a first integral that solves the Maxwell equations
\begin{align} \label{eq:sol_h}
\partial_{r}\left( \sqrt{-g}\, e^{-b\phi}F^{r\phi_i} \right)\propto\partial_{r}\left( \sqrt{-g}\, e^{-b\phi}F^{rt} \right)=0\quad \Rightarrow\quad h'(r)=\frac{ Q  }{e^{-b\phi}} \left (\frac{\ell}{r} \right)^{d-1-z},
\end{align}
where $Q$ is an arbitrary integration constant. 

In the 4-dimensional case we consider a general vector potential compatible with the spacetime symmetries, $A=A_t(r)dt+A_\phi(r)d\phi$. Thus, from Maxwell equations we are able to get two first integrals with their respective integration constants. However, the evaluation of the complete set of equations \eqref{eq:EoM} restricts these constants to be equal and we arrive to the same result coming from the ansatz \eqref{eq:FieldAnsatz}, after a constant redefinition. We expect something similar to happen in higher dimensions, being the reason to assume the vector potential proposed in \eqref{eq:FieldAnsatz}. Note that along with the electric charge there are multiple magnetic charges in our configuration as it can be seen from the Maxwell equations in \eqref{eq:sol_h}. Since all of these charges obey the same differential equation as a result of the assumed ansatz \eqref{eq:FieldAnsatz}, we denote them by the constant Q introduced in Eq. \eqref{eq:sol_h}.
  
A combination of the Einstein equations that decouple the gravitational potential is
\begin{equation}\label{eq:EtpEy}
\mathcal{E}^r_{r}+\mathcal{E}^{y_k}_{y_k}=\dfrac{1}{2\ell^2}\left[  r^2f''+\left(3(z+d)-7\right)rf'+2(z+d-2)(z+d-3)f+4 \lambda\ell^2\right]=0,  
\end{equation}
which corresponds to a second order inhomogeneous Euler equation for $f$ with the following general  solution
\begin{equation}\label{eq:MetricPot}
f(r)=\frac{k_1}{r^{z+d-2}}+\frac{k_2}{r^{2(z+d-3)}}-\dfrac{2\ell^2\lambda}{(z+d-2)(z+d-3)},
\end{equation}
where $k_1$ and $k_2$ are new independent integration constants. By imposing the Lifshitz boundary condition, $\displaystyle{\lim_{r\to\infty}f(r)=1}$, we get a negative definite cosmological constant $\lambda=-\dfrac{(z+d-2)(z+d-3)}{2\ell^2}$.

The dilaton field can be found from the following Einstein equations       
\begin{align}\label{eq:Etphi}
\mathcal{E}_t^{\phi_i} & =\dfrac{1}{2}a_i\Xi\qty{-2\kappa Q^2e^{b\phi}\qty(\dfrac{\ell}{r})^{2(d-2)}+\dfrac{1}{\ell^2}\qty[r^2f''+(3z+d-3)rf'+2(z-1)(z+d-2)f]}\nonumber\\
                       & =0,
\end{align}
by taking the gravitational potential \eqref{eq:MetricPot} as a source, obtaining
\begin{equation}
e^{b\phi}=\dfrac{1}{\kappa Q^2\ell^2}\qty[(d-2)(z+d-4)\dfrac{k_2}{r^{2(z+d-3)}}+(z-1)(z+d-2)]\qty(\dfrac{r}{\ell})^{2(d-2)}. 
\end{equation}
The remaining Einstein equations are not identically fulfilled but establish the constraints $b=2\sqrt{\dfrac{\kappa(d-2)}{z-1}}$ and $k_2=0$. Gathering all this information and renaming the constant $k_1=-M$, the solution will be given by the spacetime defined in \eqref{eq:Cov_Ansatz} with the following gravitational potential and matter fields    
\begin{subequations}
\label{eq:TaylorSol}
\begin{align}
 f(r) & =1-\dfrac{M}{r^{z+d-2}},\label{eq:TaylorSola}\\
 e^{-b\phi} & =\kappa\dfrac{Q^2\ell^2}{(z-1)(z+d-2)}\left(\dfrac{\ell}{r}\right)^{2(d-2)},\label{eq:phi}\\
 A_{t} & =-\Xi\dfrac{(z-1)}{\kappa Q \ell}\left(\dfrac{r}{\ell}\right)^{d+z-2},\\
 A_{\phi_i} & =a_i\dfrac{(z-1)}{\kappa Q \ell}\left(\dfrac{r}{\ell}\right)^{d+z-2},\\
 \lambda & =-\dfrac{(z+d-2)(z+d-3)}{2l^2},\\
 b & =2\sqrt{\dfrac{\kappa(d-2)}{z-1}},
\end{align}
\end{subequations}
where the integration constants of $A_t$ and $A_{\phi_i}$ were chosen to vanish.
%
 
We would like to mention here that the limit $z\rightarrow 1$ for the scalar field is  smooth. In fact, the apparent singularity of (\ref{eq:phi}) is due to the way in which we present the solution.

On the other hand, the limit $Q\rightarrow 0$ is not smooth. From Eqs. (\ref{eq:FieldAnsatz}) and (\ref{eq:sol_h}) one realizes that such a limit turns off the interaction between the scalar field and the Maxwell potential. This case effectively reduces the action (\ref{eq:EMD}) to the Einstein-Hilbert action with a cosmological constant plus the kinetic term for the scalar field, however, by solving the field equations one realizes there is no Lifshitz black hole configuration supported by this action. Therefore, it is necessary to appropriately modify this scalar-tensor theory (with curvature corrections –as in New Massive Gravity- or with other matter fields) to support our rotating Lifshitz black hole configurations.

Finally, the configuration (\ref{eq:TaylorSol}) corresponds indeed to the Taylor static black hole solution quoted in \cite{Taylor:2008tg} when all the rotation parameters $a_i$ are set to zero.

\section{Conserved Charges and Black Hole Thermodynamics\label{Sec:Thermodynamics}}

\subsection{Conserved charges\label{Subsec:Conserved charges}}

In order to find the energy and angular momenta associated to our rotating black hole configuration we shall make use of an off-shell formalism called generalized ADT quasi-local method, foremost introduced in 
\cite{Yi:2013}. This method has been implemented and proved to be well suited in the construction of conserved charges in higher order gravity theories (see for instance \cite{Ayon-Beato:2015jga, Gaete:2017, Ayon-Beato:2019aa}), as well as in EMD theories (\cite{Yi:2014, HerreraAguilar:2020iti}). In this method, the conserved charge corresponding to a Killing vector field $\xi$ is given by
\begin{equation} \label{eq:conserved_charge}
\mathcal{Q} (\xi) = \int d^{d-2} x_{\mu \nu} \left(\Delta K^{\mu \nu}(\xi) -2 \xi^{[\mu}\int_{0}^{1} ds \Theta^{\nu]} (\xi,s) \right),
\end{equation}
where $d^{d-2}x_{\mu \nu} = \frac{1}{2}  \frac{1}{(d-2)!} \epsilon_{\mu \nu \mu_{1} \mu_{2}\ldots \mu_{d-2}}dx^{\mu_{1}} \wedge\ldots \wedge dx^{\mu_{d-2}}$ with $\epsilon_{\mu \nu \mu_{1} \mu_{2}\ldots \mu_{d-2}}$ is the totally antisymmetric  Levi-Civita symbol, whereas $s$ stands for a parameter allowing to interpolate the black hole configuration between the solution of interest ($s=1$) and the asymptotic one  ($s=0$). Furthermore, $\Delta K^{\mu \nu} (\xi)= K^{\mu \nu}_{s=1} (\xi)-K^{\mu \nu}_{s=0} (\xi)$ stands for the total difference of the Noether potentials between the two end points of the path, $s=1$ and $s=0$.  Finally, $\Theta^{\nu}$  is the surface term obtained after the variation of the corresponding action.

The Noether potential and the surface terms associated to our model (\ref{eq:EMD}) are respectively given by 
\begin{equation} \label{eq:NoetherP}
K^{\mu \nu}(\xi) = 2 \sqrt{-g} \left[\frac{\nabla^{[\mu}\xi^{\nu]}}{2 \kappa}-\frac{1}{2}\frac{\partial \mathcal{L}}{\partial (\partial_{\mu} A_{\nu})}  \xi^{\sigma}A_{\sigma}\right],
\end{equation}
\begin{equation} \label{eq:surfaceT}
\Theta^{\mu}(\delta g, \delta \phi , \delta A ) =  2 \sqrt{-g} \left(  \frac{g^{\alpha[ \mu }\nabla^{\beta]} \delta g_{\alpha \beta}}{2 \kappa}  +\frac{1}{2} \frac{\partial \mathcal{L}}{\partial (\partial_{\mu} A_{\nu})} \delta A_{\nu} + \frac{1}{2} \frac{\partial \mathcal{L}}{\partial (\partial_{\mu} \phi)} \delta \phi\right).
\end{equation}
In order to derive the corresponding mass and angular momenta related to  our model (\ref{eq:EMD}) and black hole configuration (\ref{eq:Cov_Ansatz}) we shall consider a timelike and a set of rotational Killing vector fields, respectively. 
 
For the case of the mass, the Killing vector takes the form $\eta^{\mu}=-\partial_{t}$. This vector field, together with (\ref{eq:conserved_charge}), (\ref{eq:NoetherP}), and (\ref{eq:surfaceT}) eventually gives rise to the following expression for the mass
\begin{equation}\label{eq:Mass}
\mathcal{M}= \frac{V_{d-2}}{2 \kappa \ell^{z+[d/2]}} \left[(d-z)\Xi^{2}+(z-2)\right]r_{H}^{z+d-2},
\end{equation}
where $V_{d-2}$ represents the volume of the transverse spatial dimensions and the exponent $[x]$ denotes the integer part of $x$.
%
%
%

Furthermore, for the computation of the angular momenta we consider a set of rotational Killing vector fields that take the form $\zeta_i^{\mu} = \partial_{\phi_{i}}$,  allowing us to find that the angular momentum corresponding to the $i$-th rotation plane can be expressed as
%
%
\begin{equation} \label{eq:Ang_Momentum}
\mathcal{J}_{i} =\frac{a_{i} \Xi (d-z)  V_{d-2} r_{H}^{z+d-2} }{2 \kappa \ell^{z+\left[d/2\right]} }.
\end{equation}
On the basis of this fact we can assure that the black hole field configuration rotates and the $a_i$ are indeed rotation parameters.

The electromagnetic charge can be computed  through a Gaussian integral over a spatial hypersurface $\Sigma$ at asymptotic infinity \cite{Bravo-Gaete:2020ftn}
\begin{equation}\label{eq:ElectricCharge}
Q_{em}=\int_{\Sigma}d^{D-2}x\sqrt{|\gamma|}n^\mu u^\nu e^{-b\phi}F_{\mu\nu}=\Xi QV_{d-2},
\end{equation}
where $\gamma$ is the induced metric on $\Sigma$,  with $u$ and $n$ its timelike and spacelike normal unit vectors 
\begin{equation}\label{eq:NormalVectors}
 u=\dfrac{\sqrt{f(r)}\qty(\dfrac{r}{\ell})^{z+1}}{\sqrt{\qty(\dfrac{r}{\ell})^{2z}f(r)\qty(1-\Xi^2)+\qty(\dfrac{r}{\ell})^2\Xi^2}}\,dt, \qquad \quad n=\dfrac{1}{\sqrt{f(r)}}\,\qty(\dfrac{\ell}{r})\,dr.
\end{equation}
The electromagnetic charge is a conserved quantity derived from the formula \eqref{eq:ElectricCharge} and accounts for both the electric and the magnetic components of the field, which are coupled in the case under study. In fact, one can drop the magnetic part (as well as the rotating sector of the configuration) by setting the rotation parameters to zero, but the electric part is not allowed to vanish because it supports the whole field configuration.  

\subsection{Gravitational Thermodynamics}

A formal relation between black holes and thermodynamics was systematically presented by Bardeen, Carter and Hawking in \cite{Bardeen:1973gs}, inspired by early works of Smarr \cite{Smarr:1972kt} and Bekenstein. In \cite{Bardeen:1973gs} it is emphasized that the four laws of black hole mechanics represent an analogy more than a physical reality. The analogy was pushed even further by Bekenstein himself \cite{Bekenstein:1973ur} suggesting that the area of the event horizon and the surface gravity were indeed related to the entropy and the temperature of the black hole, respectively. Hawking further justified part of this statement assuming particle creation near the event horizon \cite{Hawking:1974sw} due to quantum mechanical effects over a curved spacetime. Part of such particles tunnels through the event horizon and part of them escape to null infinity producing a black body radiation with Hawking temperature. This temperature is defined in terms of the surface gravity $\tilde{\kappa}$ by the relation \cite{Hawking:1974sw} 
\begin{equation} \label{eq:Hawking_Temp0}
T_H \equiv \frac{\tilde{\kappa}}{2 \pi},
\end{equation}
where the surface gravity can be expressed in terms of the null generator of the event horizon $\chi^{\mu}$, according to
\begin{equation}\label{eq:Surface_g}
\tilde{\kappa}^{2}= -\frac{1}{2}  \left(\nabla_{\mu}\chi_{\nu}\right)\left(\nabla^{\mu} \chi^{\nu}\right).
\end{equation}
For our rotating black hole geometry this corresponds to the Killing vector field 
\begin{equation}\label{eq:Killing_F}
\chi^{\mu}= \partial_{t}+\sum_{i=1}^{n} \Omega_{i} \partial_{\phi_{i}},
\end{equation}
with $\Omega_{i}$ the angular velocity linked to the $i$-th rotation plane that can be obtained by considering the null character of the Killing vector field (\ref{eq:Killing_F}) at the horizon. By doing this, one can easily find that the $i$-th angular velocity related to the $i$-th rotating axis is given by
\begin{equation}\label{eq:Ang_vel}
\Omega_{i} = \frac{a_{i}}{\ell^{2} \Xi}, 
\end{equation}
from which we can express $\Xi$ in terms of $\Omega$ as follows
\begin{equation}\label{eq:Xi}
\Xi = \frac{1}{\sqrt{1-\ell^2 \Omega^2}},
\end{equation}
subject to the restriction $\Omega^2 \equiv \sum_{i} \Omega_{i}^2 < 1/\ell^2$. 

All relations in black hole Thermodynamics are about geometrical quantities \cite{Bardeen:1973gs,Hawking:1974sw,Wald:1993nt,Iyer:1994ys,Iyer:1995kg}. The holographic duality between gravity and condensed matter systems establishes a relation between the black hole Hawking temperature and the temperature of a condensed matter system in a given thermodynamic phase. This motivation allows us to realize that the Hawking temperature associated with the configuration (\ref{eq:Cov_Ansatz}) reads
\begin{equation} \label{eq:Hawking_Temp}
T_H =T-T_c \equiv \frac{\tilde{\kappa}}{2 \pi},
\end{equation}
where $T_c\ge 0$ is certain constant critical value of the temperature, since there is no restriction to map $T_H\rightarrow T-T_c$ as far as this quantity remains positive for a given thermodynamic phase. This difference allows us to consider processes approaching the critical temperature from above where the critical point is reached at the extremal case $\ell^2\Omega^2=1$, 
evidencing the emergence of a possible phase transition of the condensed matter system at finite temperature
(see equations \eqref{eq:Xi} and \eqref{eq:Temp}). Moreover, this definition allows us to give a consistent physical interpretation of the low temperature limit of the Reissner-Nordstr\"om-AdS black hole geometry, avoiding the issue of having a state with finite entropy and zero temperature that contradicts the third law of Thermodynamics \cite{Hartnoll2016vb,Sachdev2012qw}.

Alternatively, in order to take into account thermodynamic processes approaching the critical temperature from below and reaching a phase transition we can use the following definition
\begin{equation} \label{eq:Hawking_Tempinv}
T_H =T_c-T \equiv \frac{\tilde{\kappa}}{2 \pi},
\end{equation}
where now $T_c-T$ is a positive definite quantity.

Thus, this definition allows us to compute the Hawking temperature \eqref{eq:Hawking_Temp} associated with the rotating black hole field configuration (\ref{eq:Cov_Ansatz}): 
\begin{equation}\label{eq:Temp}
T_{H} = T-T_c = \frac{1}{4 \pi} \frac{(z+d-2) r_{H}^z}{\ell^{z+1} \Xi},
\end{equation}
whereas the entropy, giving in terms of the event horizon area  $\mathcal{A}$, adopts the form
\begin{equation}\label{eq:Entropy}
\mathcal{S} \equiv \frac{\mathcal{A}}{4 G_{d}} = \frac{\Xi V_{d-2}}{4 G_{d} \ell^{[d/2]-1}} r_{H}^{d-2},
\end{equation}
where $G_{d}=\kappa/8\pi$ is the $d$-dimensional gravitational constant.

Finally, by making use of the expressions (\ref{eq:Mass}), (\ref{eq:Ang_Momentum}), \eqref{eq:Ang_vel}, (\ref{eq:Temp}) and  (\ref{eq:Entropy}), it is easy to check that the first law of thermodynamics is reasonably satisfied 
\begin{equation}\label{eq:FLaw}
d \mathcal{M} = \left(T-T_c\right) d \mathcal{S}+\sum_{i=1}^{n} \Omega_{i} d \mathcal{J}_{i}+\Phi_{em}dQ_{em}, 
\end{equation}
with the electromagnetic potential
\begin{equation}
 \Phi_{em}=-\dfrac{(z-1)}{\kappa Q\ell^{z+[d/2]}}\dfrac{r_h^{z+d-2}}{\Xi},
\end{equation}
obtained as an appropriate linear combination of $A_t$ and $A_{\phi_i}$ at the event horizon. The variations in the first law of black hole Thermodynamics \eqref{eq:FLaw} are taken with respect to two sets of independent variables, for instance the temperature $T$ and the angular velocities $\Omega_i$. In fact, the electromagnetic charge contributes to the variations only with respect to the angular velocities.\\
The Smarr-like formula adopts the following form
\begin{equation}\label{eq:SmarrFormula}
\mathcal{M} = \left[\dfrac{d-2+2\,c\,(z-1)}{z+d-2}\right]\left(T-T_c\right) \mathcal{S}+\sum_{i=1}^{n} \Omega_{i}\mathcal{J}_{i}+c\,\Phi_{em}Q_{em}, 
\end{equation}
where $c$ is an arbitrary constant.

These remarkable relations provide striking evidence that the Thermodynamics of our rotating black hole configuration is physically consistent.

\section{A look for the thermodynamic stability\label{Sec:NTG}}
Here we explore the thermodynamic stability of the constructed solution from a local and a global perspective. We study local stability by computing the associated specific heats, while the global one is explored using the so-called new thermodynamic geometry (NTG) formalism as appears in \cite{Hosseini:2019}. It is worth mentioning that the NTG formalism corresponds to one of the most recent attempts for describing the thermodynamics of black holes using Riemannian geometry. Although there are several earlier developments \cite{Weinhold:1975,Ruppeiner:1995zz,Quevedo:2006xk,Bravetti_2013,Quevedo:2019wbz}, this formalism has proved to be the most suitable one since, besides being Legendre invariant, it also solves some of the inconsistencies presented in the aforementioned developments. 

Before we proceed, we remark that it is convenient to work in the Gibbs free energy representation since this potential allows us to construct a positive definite metric \cite{Kolev:2019tqa}. In this sense, the Gibbs free energy can be obtained from a Legendre transformation of the energy \eqref{eq:SmarrFormula}, resulting in
\begin{align}
\mathcal{G}(T, \Omega_{1},\ldots, \Omega_{n}) &=  \mathcal{M}-\qty(T-T_c) \mathcal{S}-\sum_{i=1}^{n} \Omega_{i}\mathcal{J}_{i},\nonumber \\
  &= -\frac{z V_{d-2}}{2 \kappa \ell^{z+[d/2]}} \qty[\frac{4\pi \qty(T-T_c) \ell^{z+1} \Xi}{z+d-2}]^{\frac{z+d-2}{z}},  
  \label{eq:GibbsPot}          
\end{align}
in differential form this relation reads
\begin{align}\label{eq:dGibbsPot}
d\mathcal{G}(T,\Omega_{1},\ldots, \Omega_{n}) & = -\mathcal{S} dT-\sum_{i=1}^{n} \mathcal{J}_{i}d\Omega_{i}+\Phi_{em}dQ_{em},\nonumber\\
& = -\mathcal{S} dT-\sum_{i=1}^{n} \left(\mathcal{J}_{i}-\Phi_{em}\dfrac{\partial Q_{em}}{\partial\Omega_i}\right)d\Omega_{i},
\end{align}
where in the last equality we have taken into account the dependence of the electromagnetic charge on the angular velocities \eqref{eq:ElectricCharge}.
 
The angular momenta and entropy written in terms of the Gibbs free energy natural variables $(T, \Omega_{1},\ldots, \Omega_{n} )$, are given respectively by
\begin{equation}\label{eq:am_e}
\mathcal{J}_{i}= \frac{\alpha z (d-z)}{z+d-2} \ell^2 \Omega_{i} \Xi^{\frac{3z+d-2}{z}}\left(T-T_c\right)^{\frac{z+d-2}{z}}, \qquad \mathcal{S} = \alpha z \Xi^{\frac{z+d-2}{z}}\left(T-T_c\right)^{\frac{d-2}{z}}
\end{equation}
where for simplicity we have defined $\alpha = \frac{(z+d-2)}{z} \frac{V_{d-2}}{2\kappa \ell^{z+[d/2]}}\qty(\frac{4\pi \ell^{z+1}}{z+d-2})^{\frac{z+d-2}{z}}$. Additionally, in these expressions we have made use of Eq.  \eqref{eq:Xi}.

As it was previously stated, we shall make use of the specific heats in order to study the local thermodynamic stability and phase transitions of the system. In the $(T,\Omega_1\ldots,\Omega_n)$ space of the Gibbs free energy, we define the specific heats of the black hole following \cite{Mansoori:2015} as follows
\begin{equation}\label{eq:spec_heat_0}
\begin{split}
C_{\Omega_1,\ldots,\Omega_n} &= \left(T-T_c\right) \qty(\pdv{\mathcal{S}}{T})_{\Omega_1,\ldots, \Omega_n}=  \left(T-T_c\right)  \frac{\{\mathcal{S}, \Omega_1,\ldots,\Omega_n\}_{T,\Omega_1,\ldots,\Omega_n}}{\{T,\Omega_1,\ldots,\Omega_n\}_{T,\Omega_1,\ldots,\Omega_n}}\\
&= \alpha (d-2) \Xi^{\frac{d+z-2}{z}} \left(T-T_c\right)^{\frac{d-2}{z}},
\end{split}
\end{equation}
and
\begin{equation}\label{eq:spec_heat_1}
\begin{split}
C_{\mathcal{J}_1,\ldots,\mathcal{J}_n} &= \left(T-T_c\right) \qty(\pdv{\mathcal{S}}{T})_{\mathcal{J}_1,\ldots, \mathcal{J}_n}=  \left(T-T_c\right)  \frac{\{\mathcal{S}, \mathcal{J}_1,\ldots,\mathcal{J}_n\}_{T,\Omega_1,\ldots,\Omega_n}}{\{T,\mathcal{J}_1,\ldots,\mathcal{J}_n\}_{T,\Omega_1,\ldots,\Omega_n}}\\
&= -\frac{\alpha z \Bigl[z \ell^2 \Omega^2 -(d-2)\Bigr]}{\ell^2 \Omega^2(d+2z-2)+z} \Xi^{\frac{d+z-2}{z}} \left(T-T_c\right)^{\frac{d-2}{z}}.
\end{split}
\end{equation}
From these relations we observe that both specific heats are regular everywhere except when $\Omega^2 =1/\ell^2$, where $\Xi$ is singular, evidencing the emergence of a possible phase transition in the thermodynamic system. This extremal point corresponds to one in which the metric is not well-defined. Moreover, we also identify a change of sign occurring in $C_{\mathcal{J}_1,\ldots,\mathcal{J}_n}$ when 
\begin{equation}\label{eq:sign_change}
\ell^2 \Omega^2 = \frac{d-2}{z},
\end{equation}
pointing to the existence of a thermodynamic instability.

It is well-known that in order to have local thermodynamic stability in a system, one must require both specific heats to be positive definite. This fact is fulfilled by imposing the condition 
\begin{equation}\label{eq:stability_req}
\ell^2 \Omega^2 < \frac{d-2}{z}. 
\end{equation} 
On the other hand, we can also consider a weaker stability condition such that $\ell^2 \Omega^{2}<1$, implying the black hole being locally stable concerning $C_{\Omega_1, \ldots, \Omega_n}$, but allowing the system to suffer a phase transition with respect to $C_{\mathcal{J}_1, \ldots, \mathcal{J}_n}$ from a stable into an unstable one due to a change of sign.

An alternative way for exploring the stability but now from a global perspective is through the NTG formalism, as we remarked earlier. In this formalism, the thermodynamic metric in the  Gibbs free energy representation is defined according to the relation 
\begin{equation}\label{eq:Tmetric}
g_{ab} = -\frac{1}{\left(T-T_c\right)} \partial_{a} \partial_{b} \mathcal{G}(T,\Omega_{i}),
\end{equation}
with the indices $(a,b)$ taking the values $a,b = T, \Omega_{1},\ldots, \Omega_{n}$. The explicit form of the metric components read
\begin{subequations}\label{eq:Tmetric_comp}
\begin{align}
g_{TT} & =\alpha (d-2) \Xi^{\frac{d+z-2}{z}}\left(T-T_c\right)^{\frac{d-2z-2}{z}},\\
g_{T\Omega_{j}} & =\alpha (z+d-2) \ell^{2} \Xi^{\frac{d+3z-2}{z}}\left(T-T_c\right)^{\frac{d-z-2}{z}} \Omega_{j}, \\
g_{\Omega_i \Omega_j} & =\alpha \ell^{2} \left\{(3z+d-2)\ell^{2}\Omega_{i} \Omega_{j}  \Xi^{3}+z\delta_{ij}\right\}\Xi^{\frac{d+3z-2}{z}}\left(T-T_c\right)^{\frac{d-2}{z}}.
\end{align}
\end{subequations}
In addition we can write the metric in a differential form as follows
\begin{equation}\label{eq:dTmetric}
ds_{\mathcal{G}}^2 = g_{TT} dT^2 +\sum_{ij} g_{\Omega_{i} \Omega_{j}} d\Omega_{i} d\Omega_{j} +2 \sum_{i}g_{T \Omega_{i}}dT d\Omega_{i},
\end{equation}
from which we find that the curvature scalar is given by
\begin{equation}\label{eq:Tscalar}
\begin{split}
R_\mathcal{G} =-\dfrac{\qty(1-\ell^2 \Omega^2)^{\frac{z+d-2}{2z}}(n-1)}{\alpha z^2 \qty[z\ell^2 \Omega^2-(d-2)]^2\left(T-T_c\right)^{\frac{d-2}{z}}} & \Biggl\{(n-2)z^3\left[\ell^2 \Omega^2-1\right]\ell^2 \Omega^2+ \\ &\frac{n(d-2)}{4}\Biggl[ ( z \ell^2 \Omega^2 -2z-d+2)^{2}-4z^2 \ell^2 \Omega^2\Biggr]\Biggr\}.
\end{split}
\end{equation}
From this expression, it is evident that it vanishes when $d=4\to n=1$ and, for $d>4$, we observe that the scalar is negative definite. Besides, we identify the appearance of two singularities: one takes place when the temperature adopts some critical value $T=T_c$ and another one when the angular velocities are restricted by $\ell^2 \Omega^2 =(d-2)/z<1$. It is interesting to remark that the singularity at  $T=T_c$ does not appear in the Gibbs potential or the specific heats, but it is present in the second derivative of the Gibbs potential and the first derivative of the specific heats, provided $(d-2)/z <1$. Therefore, we can say that a second-order phase transition takes place due to thermal fluctuations.

On the other hand, the singularity at $\ell^2 \Omega^2 =(d-2)/z<1$ does not appear in either the Gibbs potential or the specific heats, but it does coincide with the change of sign in $C_{\mathcal{J}_1,\ldots, \mathcal{J}_n}$. As we have remarked earlier, a transition from a stable into an unstable black hole occurs at this singularity. 

It is interesting to note that the metric singularity at the extremal point $\ell^2 \Omega^2 =1$ is not present in the scalar curvature, which vanishes at this point, clarifying that it is in fact a coordinate singularity. On the other hand, the specific heats diverge at this point as mentioned earlier. A similar situation takes place in the thermodynamic system studied in \cite{Kolev:2019tqa}, where by analyzing the specific heats in the entropic representation one sees that this divergence does not appear, confirming the coordinate nature of the aforementioned singularity in these physical quantities.

Finally, we remark that these results are in perfect accordance with the prescription of thermodynamic geometry, in the sense that singularities of the curvature scalar for the thermodynamic metric coincide with the phase transitions observed in the specific heats \cite{Mansoori:2014}.

\section{Causal Structure\label{Sec:Physics}}
\begin{center}
\begin{figure}[ht!]\centering
  \includegraphics[width=0.58\linewidth]{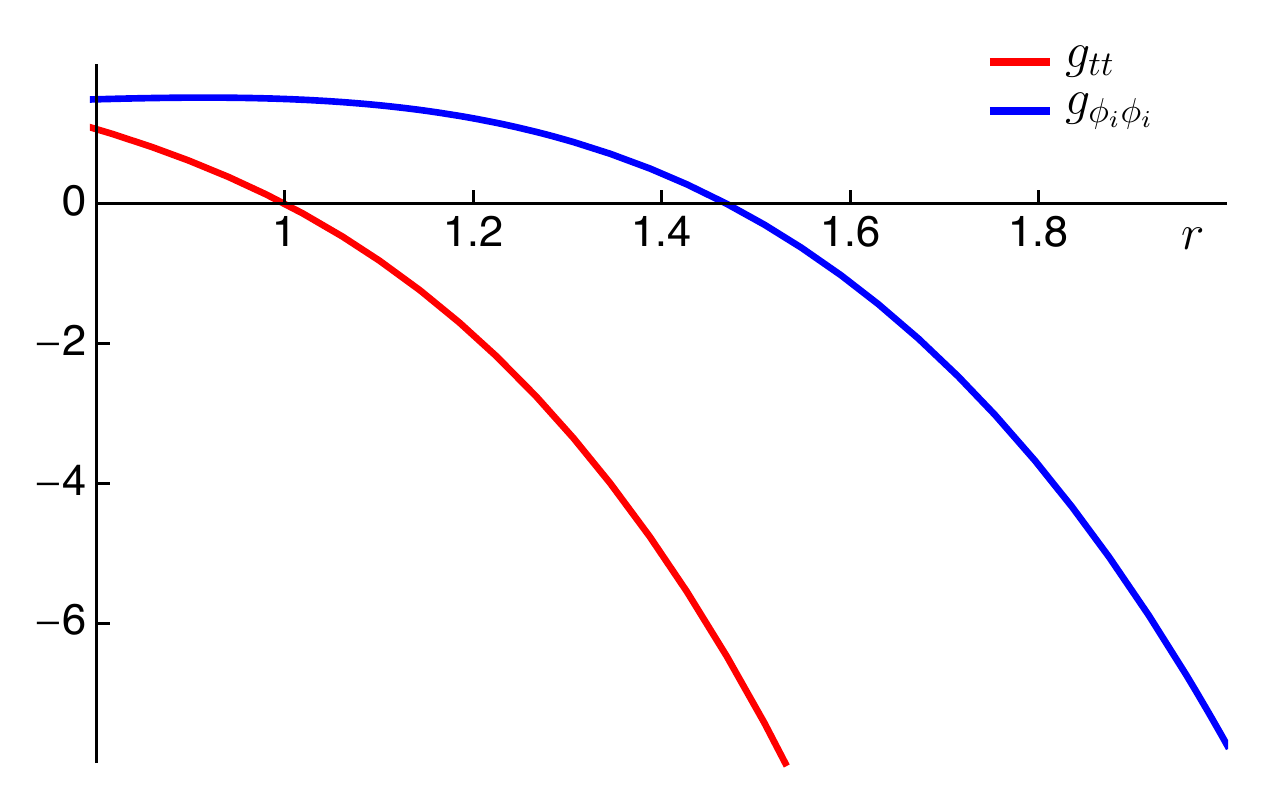}
  \caption{The metric components $g_{tt}$ and $g_{\phi_i\phi_i}$ with parameters $\ell=1$, $M=0.5$, $a_i=1$, $z=2$ and $d=5$. The vertical axis signals the event horizon radius $r_H$. From the event horizon to the intersection of the \emph{red} curve with the r-axis we have the \emph{ergosphere}. From the intersection of the \emph{blue} curve with the r-axis to infinity we have \emph{closed timelike} curves.}
  \label{fig:Causal}
\end{figure}
\end{center}

The anisotropic nature of the metric \eqref{eq:Cov_Ansatz} makes the spacetime structure more involved with respect to the (A)dS rotating spacetimes \cite{Awad:2002cz}. For example, in this black hole field configuration the \emph{ergosphere} is defined as the region where the norm of the timelike Killing field $\eta=\partial_t$ becomes positive
\begin{align}\label{eq:Ergosphere}
 \eta^2 & =\eta^\mu g_{\mu\nu}\eta^\nu\equiv g_{tt},\nonumber\\
        & =\qty[-\qty(\dfrac{r}{\ell})^{2z}f(r)+\qty(\dfrac{r}{\ell})^2]\Xi^2-\qty(\dfrac{r}{\ell})^2\nonumber\\
        & =-\qty(\dfrac{r}{\ell})^{2z}f(r)+\qty(\dfrac{r}{\ell})^2\sum_{i=1}^{n}\qty(\dfrac{a_i}{\ell})^2>0.
\end{align}
Below the radius of the event horizon $f(r)<0$ and the inequality \eqref{eq:Ergosphere} is satisfied. However, a change of sign takes place at some point outside the event horizon as it will be exemplified in Figure~\ref{fig:Causal} for some parameters election. Hence, there is a region between $r_H$ and the largest root of $\eta^2=0$ where an observer cannot remain stationary. The spacetime drags the observer even when he is outside the black hole's event horizon.

Another issue that arises within this metric is related to the rotation Killing fields $m_i$. There are $n$ independent rotation planes and over each plane there is a rotation Killing vector field $m_i=\partial_{\phi_i}\equiv m_i^\mu\partial_\mu$, $i=1\hdots n$, which is cyclic -- $\phi_i\in[0,2\pi]$ --. Their norm reads
\begin{align}\label{eq:NormCyclic}
m_i^2 & = m_i^\mu g_{\mu\nu}m_i^\nu\equiv g_{\phi_i\phi_i},\nonumber\\
      & = \qty[-\qty(\dfrac{r}{\ell})^{2z}f(r)+\qty(\dfrac{r}{\ell})^2]a_i^2+r^2,
\end{align}
that could be null, spacelike or timelike depending on the function $f(r)$. Let us explore where the vector is timelike, $m_i^2<0$. From \eqref{eq:NormCyclic}, by taking into account that $a_i\neq 0$, we obtain
\begin{equation}
f(r)>\qty(\dfrac{\ell}{r})^{2(z-1)}\dfrac{\qty(a_i^2+\ell^2)}{a_i^2}. 
\end{equation}
It will be timelike if $f(r)$ is greater than certain positive number and the limit $r\rightarrow\infty$ implies $f(r)>0$. In the case of our solution \eqref{eq:TaylorSola}, the function $f(r)$ is positive outside the event horizon and we will end up with closed timelike curves accessible from null infinity. This constitutes a pathology of the spacetime. 
We have already proved the spinning black hole configuration given by \eqref{eq:Cov_Ansatz} and \eqref{eq:TaylorSola} as a proper solution to the Einstein-Maxwell-Dilaton Theory \eqref{eq:EMD} with a well-defined Thermodynamics provided in Sec.~\ref{Sec:Thermodynamics}. This solution does not seem to capture this pathological behavior. It would be interesting to explore whether this pathology has any consequence in the dual field theory since so far we are not aware of any restriction of this kind on it.         

\section{Curvature invariants\label{Curvature invariants}}

Here we compute the curvature invariants of our rotating black hole spacetime in order to study its behavior.
Thus, we proceed to display the curvature Ricci scalar 
\begin{align}\label{eq:R}
R = -\frac{1}{\ell^2}\qty{2\qty[z^2+(d-2)z+\frac{1}{2} (d^2-3d+2)]+\frac{(d-2)(z-1) M}{r^{z+d-2}}},
\end{align}
the square contraction of the Ricci tensor $\mathcal{R} =R_{\mu\nu}R^{\mu\nu}$
\begin{align}\label{eq:Ricci2}
\mathcal{R} =& \frac{1}{\ell^4}\Biggl\{2(z^2+d-2)\qty[z^2+(d-2)z+\frac{1}{2}(d^2-3d+2)]+\frac{2(d-2)(z-1)(z^2+d-2)}{r^{z+d-2}}\nonumber\\
 &+\frac{(z-1)^2 (d-2)^2 M^2}{r^{2(z+d-2})}\Biggr\},
\end{align}
as well as the Kretschmann invariant --- defined as the square contraction of the Riemann tensor $\mathcal{K}=R^{\mu\nu\alpha\beta}R_{\mu\nu\alpha\beta}$ ---
\begin{align}\label{eq:Kretschmann}
\mathcal{K}= &\dfrac{d-2}{\ell^4}\Biggl\{ [d^3-2(z+2)d^2+(z^2+8z+2)d-6(2z-1)]\qty(\dfrac{M}{r^{z+d-2}})^2\nonumber\\
& -4(z-1)\qty[zd-(z+1)^2]\qty(\dfrac{M}{r^{z+d-2}})+\dfrac{2d\qty(d+2z^2-3)+4(z^2-1)^2}{d-2}\Biggl\}.
\end{align}
All of these invariants are singular at $r=0$ and regular at the black hole horizon, $f(r_H)=0$, for all $z\geq1$ and $d>2$. Moreover, these scalars tend to a constant at spatial infinity.

Here some comments are in order regarding the structure of these curvature invariants. Their structure is the same for both static and rotating black holes due to the metric ansatz \eqref{eq:Cov_Ansatz} and the flat topology of the event horizon that have been employed to construct our stationary black hole configuration. In fact we have a rotating planar brane, therefore rotation is compatible with its planar symmetry. Precisely the same situation takes place for the famous AdS$_3$ BTZ black hole \cite{Banados:1992wn} as well as for the Lemos black hole \cite{Lemos:1994xp} and its multi-dimensional version given by the Awad metric \cite{Awad:2002cz} (all these black holes are rotating and possess non-trivial angular momenta according to the computation of the corresponding conserved global charges). When computing these curvature invariants, a peculiar mathematical fact arises indicating that they coincide for both static and rotating black holes as it can be straightforwardly appreciated from their expressions: they do not depend on the rotation parameters of the black hole solution or the transverse coordinates of the line-element since the latter are cyclic due to the flat topology they represent (this is in contrast to the spherical or hyperbolic symmetries where these coordinates are not cyclic anymore). 

At least for the case of multi-dimensional ($d\ge 4$) rotating black hole spacetimes, either in AdS or Lifshitz backgrounds, which are very similar since both possess negative curvature, one would expect more complex metrics to exist in the sense that the blackening factor entering the line-element can have a functional dependence on the angular momenta of the metric, rendering curvature invariants that explicitly depend on the rotating parameters. This is an interesting issue which is currently under research within our group.

\section{Conclusions\label{Sec:Conclusions}}
We have presented a rotating black hole exact solution, given by Eqs. \eqref{eq:Cov_Ansatz} and \eqref{eq:TaylorSola}, within the context of Einstein-Maxwell-Dilaton theory. We were able to prove that the configuration rotates with angular momenta given by Eq. \eqref{eq:Ang_Momentum}, where the $a_i$-constants can be identified as rotation parameters. Once the rotation parameters are set to zero, an asymptotically Lifshitz black hole is recovered. In addition, we implemented the off-shell quasi-local analysis for computing the mass of the black hole configuration \eqref{eq:Mass} and compute the entropy \eqref{eq:Entropy} to verify that the first law of black hole mechanics \eqref{eq:FLaw} is plausible satisfied. In order to achieve this goal we introduced a generalization of the Hawking temperature notion \eqref{eq:Hawking_Temp}, where the critical temperature can attain any positive definite or vanishing value, allowing us to easily identify the emergence of phase transitions at finite temperature and entropy, in accordance with the third law of Thermodynamics and the phenomenology of many quantum matter systems showing superconductivity or superfluidity, for instance.
Moreover, a Smarr-like formula for quasi-homogeneous functions has been constructed to give the relation \eqref{eq:SmarrFormula}. By making use of the thermodynamic quantities of our spinning black hole configuration we constructed a consistent geometrization on the basis of the Gibbs free energy. We revealed the existence of critical points where phase transitions arise in the specific heats, while singularities emerge in the curvature invariants of the thermodynamic metric. We find a remarkable correspondence between them, in complete concordance with the prescriptions of thermodynamic geometry. 
We also explored a pathological behavior of the spacetimes defined by \eqref{eq:Cov_Ansatz} and \eqref{eq:TaylorSola}, that is not captured by the Thermodynamics of the constructed metric. It would be interesting to explore whether there is any sensible consequence of this pathology in the dual field theory within the framework of the holographic correspondence. 
We finally computed and analyzed the structure of the curvature invariants of our rotating black hole configuration.     
  
\acknowledgments
The authors would like to thank Eloy Ay\'on-Beato, Mokhtar Hassa\"ine, H\'ector I. Zapata Rodr\'iguez and Manuel de la Cruz L\'opez for multiple enlightening and fruitful discussions. We are also indebted to an anonymous referee for constructive reports that helped improving our work. A.H.-A. and D.F.H.-B. thank the Sistema Nacional de Investigadores (SNI) for support. The authors acknowledge financial support from a CONACYT grant No. A1-S-38041. D.F.H.-B is also grateful to CONACYT for a Postdoc for Mexico grant No. 372516. J.A. H.-M. also acknowledges support from CONACYT through a PhD grant No. 750974.  
\bibliography{Bibliography}

\end{document}